\documentclass{aa}

\usepackage{graphicx}
\usepackage{amsfonts}
\usepackage{amsmath}
\usepackage{longtable}
\usepackage{txfonts}

\newcommand{\msun}{M_{\odot}}

\graphicspath{{/home/estelle/NBody/NBody3/RUNS_wBD/}}

\begin{document}

  \title{$\eta$ Cha: abnormal IMF or dynamical evolution~?}

  \author{E. Moraux\inst{1}\and W. A. Lawson\inst{2}\and
  C. Clarke\inst{3}}

  \offprints{Estelle.Moraux@obs.ujf-grenoble.fr}


  \institute{Laboratoire d'Astrophysique, Observatoire de Grenoble
    (LAOG), B.P. 53, 38041 Grenoble Cedex 9, France
    \and School of Physical, Environmental and Mathematical Sciences,
    University of New South Wales, Australian Defence Force Academy, 
    Canberra ACT 2600, Australia
    \and Institute of Astronomy, University of Cambridge, Madingley
    Road, Cambridge. CB3 0HA, UK
    }

  \date{Received 30/11/06; Accepted 13/07/07}

  \authorrunning{E. Moraux et al.}

  \titlerunning{}

  \abstract {$\eta$ Chamaeleontis is a unique young ($\sim9$ Myr)
  association with 18 systems concentrated in a radius of
  $\approx35$ arcmin, i.e. 1pc at the cluster distance of 97pc. No
  other members have been found up to 1.5 degrees from the cluster
  centre. The cluster mass function is consistent with the IMF of
  other rich young open clusters in the higher mass range but shows a
  clear deficit of low mass stars and brown dwarfs with no objects
  below $0.1\msun$.} {The aim of this paper is to test whether this
  peculiar mass function could result from dynamical evolution despite
  the young age of the cluster.} {We performed N-body numerical
  calculations starting with a log-normal IMF and different initial
  conditions in terms of number of systems and cluster radius using
  the code NBODY3. We simulated the cluster dynamical evolution over
  10 Myr and compared the results to the observations.} {We found that
  it is possible to reproduce $\eta$ Cha when starting with a very
  compact configuration (with $N_{init}=40$ and $R_0=0.005$pc) which
  suggests that the IMF of the association might not be abnormal. The
  high initial density might also explain the deficit of wide binaries
  that is observed in the cluster.} {}


  \maketitle


\section{Introduction}

The recently-discovered $\eta$ Chamaeleontis stellar cluster (Mamajek
et al. 1999) is remarkable in several respects. It is a compact
(extent $\sim1$pc) and sparse group of 18 pre-main sequence (PMS)
systems, and it is one of the nearest clusters to the Sun with an {\it
Hipparcos} distance $d\approx97$pc. The stellar members present a high
degree of coevality with an age of $8-10$ Myr (Lawson \& Feigelson
2001, Luhman \& Steeghs 2004, Lyo et al. 2004a) and were probably born
near the Lower Centaurus Crux subgroup of the Sco-Cen OB association
(Mamajek et al. 2000, Jilinski et al. 2005). Despite the youth of the
cluster there is no evidence for extinction due to remnant molecular
material, and the proximity of the cluster ensures that foreground
reddening is negligible (Lyo et al. 2004a). Because of these
characterictics, the $\eta$ Cha cluster is an ideal target to study
the properties of ``older'' T-Tauri stars, such as circumstellar disk
evolution (Bouwman et al. 2006) and binarity (Lyo et al. 2004b,
Brandeker et al. 2006), and to search for its brown dwarf population.

Several photometric surveys have been performed during the last few
years with the aim of obtaining a complete census of cluster members
beyond the original X-ray-selected population. To date, 18 systems
have been discovered with a mass range of $0.15\msun$ to $3.8\msun$
(Mamajek et al. 1999, Lawson et al. 2002, Lyo et al. 2004b, Song et
al. 2004). Lyo et al. (2004b) found that the cluster mass function is
consistent with the IMF of rich young open clusters and field stars in
this mass domain. By extrapolating the mass distribution of the
Trapezium cluster (Muench et al. 2002) to lower masses, they predicted
$10-14$ additional low mass stars with $0.08\msun\le m\le 0.15\msun$
and about 15 brown dwarfs should accompany the documented stellar
population. However, Luhman (2004) examined the region within a radius
of $1.5\degr$ surrounding $\eta$ Cha using 2MASS and DENIS photometric
data and found {\it no} new members within the mass range
$0.025-0.10\msun$. Lyo et al. (2006) performed a deeper survey of the
central core region down to $\sim 13 M_{Jup}$ and found no new low
mass members either. Whether this clear deficit of very low mass stars
and brown dwarfs in the cluster mass function results from an abnormal
IMF or is due to dynamical evolution is not completely clear
yet. Although $\eta$ Cha is young and sparse, there is some hint that
dynamical evolution has already occurred. In particular, the radial
distribution of the cluster members indicates that significant mass
segregation is present with more than 50\% of the mass residing in the
inner 6 arcmin or 0.17 pc (Lyo et al. 2004b). This is the sparsest
stellar cluster for which mass segregation has been seen. Moreover,
the maximum stellar mass -- cluster mass relation proposed by Weidner
\& Kroupa (2006) yields a mass of $30-40\msun$ for the embedded
cluster at the origin of $\eta$ Cha. The actual cluster mass is $18
\msun$, which suggests that a significant fraction of the primordial
stellar members has already been lost, probably as a consequence of
gas expulsion.

Binarity studies indicate that the 18 stellar systems of $\eta$ Cha
include five confirmed (RECX 1, RECX 7, RECX 9, RECX 12, and RS Cha
which may be a triple) and three probable (ECHA J0836.2-7908, ECHA
J0838.9-7916, and $\eta$ Cha) binaries; see Table 2 of Lyo et
al. (2004b).  This yields a multiple fraction of $28-44$\% which is
$2-4$ times higher than that of solar-type field stars, although it is
of the same order as observed in other young, nearby associations such
as Taurus or TW Hydrae. Speckle observations (K\"ohler \&
Petr-Gotzens 2002) and adaptive optics imaging (Brandeker et al. 2006)
resolved only two binaries, RECX 1 and RECX 9. In both cases the
projected separation is $\sim 0.2$'' which corresponds to about 20 AU
at a distance of 97 pc, and the other binaries are not spatially
resolved. RECX 7 is a dual-lined spectroscopic binary (Lyo et
al. 2003) and RS Cha AB is an eclipsing system (Andersen 1991, Mamajek
et al. 2000). No companion with projected separation larger than 20 AU
has been found around the 18 cluster members. Brandeker et al. (2006)
computed a companion probability based on their contrast sensitivity
limit and found an upper limit of about 18\% for wide ($>30$ AU)
binaries in $\eta$ Cha. This contrasts to the $\sim$58\% in the TW
Hydrae association that is of similar age. Bouwman et al. (2006)
presents the most recent summary of binary properties for the cluster.

These properties -- the deficit of very low mass objects, mass
segregation, and the lack of wide binaries -- are rather unusual
considering the youth and sparsity of the $\eta$ Cha cluster. Whether
they were imprinted during the formation phase as a result of
different initial conditions, or whether they are due to dynamical
evolution is an open issue.

The aim of this paper is to test this second possibility. We try to
reproduce the observed $\eta$ Cha properties using N-body numerical
simulations of the cluster dynamical evolution, starting with standard
initial conditions. In the next section, we explain the model we
used for the calculations. The results and analysis are presented in
Section 3, and the conclusion is given in Section 4.


\section{Numerical simulations}


\subsection{IMF}

There is growing evidence that the IMF is independent of initial
conditions, from the stellar domain down to the substellar regime
(e.g. Kroupa 2001, Chabrier 2003, Moraux et al. 2007). It would
therefore be rather surprising to find such a different initial mass
spectrum for $\eta$ Cha with no members with masses below $0.15\msun$.
Following the log-normal IMF from Chabrier (2003)
\begin{equation*}
\xi(m) \propto \exp \left(-\frac{(\log m - \log m_0)^2}{2\,
\sigma^2}\right)
\end{equation*}
where $m_0=0.25\msun$ and $\sigma=0.52$, we estimate the number of
systems necessary to reproduce the four objects with a mass between 1
and $4\msun$ observed in the cluster\footnote{The stars/systems with
$m\ge 1\msun$ are $\eta$ Cha, the RS Cha binary, HD 75505, and at
least one of the three K-stars, RECX 1, 7 and 11.  In the absence of
dynamical masses for any of the K-type stars, their adopted masses are
somewhat model-dependent.}. We found that about 50 systems with masses
between 0.01 and $4\msun$ are mandatory. We started a first set of
numerical calculations with $N_{init}=50$ and different values of
initial cluster radius $R_0$. Then we fixed $R_0$ around the values
that give the best results in reproducing the observations and we
performed a second set of simulations with $N_{init}=30,40,60,70$ to
test the effect of the initial number of systems on the results.

\subsection{Initial cluster radius}

Recently, Adams et al. (2006) found a correlation between the number
of stars in a star forming region and the core radius $R_{c}$ using a
compilation of data on embedded clusters from Lada \& Lada (2003) and
Carpenter (2000). Clusters identified by {\it Spitzer} using
infrared-excess sources also follow this relationship (Allen et
al. 2006). This corresponds to an average surface density of cluster
members $N/R_{c}^{2}$ that is nearly constant and gives
$R_{c}=0.3-1.0$ pc for $N=50$ stars. 

In the calculations, the cluster size is defined by the virial radius
\begin{equation*}
R_0= - \frac{GM_c^2}{4E}
\end{equation*}
where $M_c$ is the total mass of the cluster, and $E$ is the total
energy of the system. To maximize the density, and thus dynamical
interactions, between cluster members, we first adopted the value
$R_0=0.3$ pc as an initial condition. As $\eta$ Cha appears to be
already evolved dynamically despite its youth -- all the low mass
objects with $m<0.1\msun$ have already been lost -- we also started
our calculations with more compact configurations down to $R_0=0.005$
pc. We chose $R_0=0.3$, 0.1, 0.05, 0.03, 0.01 and 0.005 pc.

\subsection{Initial density distribution}

Significant mass segregation is already present in $\eta$ Cha with
more than 50\% of the stellar mass residing in the inner 0.17 pc (Lyo
et al.  2004b). Similarly, studies of young rich stellar clusters
(e.g. in the ONC; Hillenbrand \& Hartmann 1998) have shown that the
most massive stars are concentrated in the central core of the
cluster. Bonnell \& Davies (1998) have suggested that mass segregation
may reflect the initial condition of the cluster, and is not a
consequence of dynamical interactions. To test this hypothesis, we
started some of our simulations with primordial mass
segregation, and some without.

In all cases, we initially select stellar positions so as to
follow a Plummer model
\begin{equation*}
\rho(r)= \frac{3 N_{init}}{4\pi R_{pl}^3}
\frac{1}{[1+(r/R_{pl})^2]^{5/2}}
\end{equation*}
with $R_{pl}=(3\pi/16)R_0$ (Aarseth et al. 1974). To add primordial
mass segregation, we simply rejected stars with $m>0.5\msun$ that were
assigned to $r>0.1R_0$ and repeated the random drawing for this
cluster member. The velocities are scaled so as to ensure global
virial equilibrium, although we note that in the mass segregated case
the initial conditions do not correspond to a solution of the steady
state collisionless Boltzmann equation. We note that rapid collisional
evolution of this small N system renders the question of whether the
initial conditions are in precise equilibrium of little importance.

\subsection{Initial speeds}

Initially the cluster is assumed to be in virial equilibrium, which is
characterized by the ratio of kinetic energy to potential energy
$Q=0.5$. The velocity distribution is everywhere isotropic, and the
generating function corresponding to a Plummer model is given by
Aarseth et al. (1974).

\subsection{Gas}

Although the details of gas removal are not fully understood,
observations indicate that clusters older than 5 Myr are rarely
associated with molecular gas. This is the case of $\eta$ Cha. Mamajek
et al. (2000) suggested that the superbubble formed by Sco-Cen OB
winds and supernova remnants terminated star formation in the cluster
and dispersed its natal molecular gas. This could have occured as
recently as only a few Myr ago.

The gas is usually represented as an external potential in cluster
evolution models, and is removed a few Myr after the star formation
process has begun on a timescale of the order of a crossing time or
less (e.g. Geyer \& Burkert 2001; Kroupa et al. 2001; Kroupa \&
Bouvier 2003; Bastian \& Goodwin 2006). At the end of the gas
expulsion, the cluster is out of virial equilibrium and the subsequent
evolution is a relaxation of the cluster as it attempts to return to
virial equilibrium. This phase is found to last for about 20 initial
crossing times $t_{cr}$ (Geyer \& Burkert 2001; Bastian \& Goodwin
2006) but we expect it to be short ($\sim 0.1-1$ Myr) in the case of
$\eta$ Cha. In order to reproduce the observations, the initial
configuration should be very compact, which suggests that $t_{cr}$
will be very small initially.

In this study, we consider that the gas has already been removed and
that the cluster has returned to virial equilibrium when we start our
calculations. This means we start the models when $\eta$ Cha is
already a few Myr old. The dynamical evolution of the cluster is then
purely due to two-body interactions and that is simulated by our
N-body numerical calculations. Therefore, the models do not have to
run for 10 Myr (the maximum estimated age of $\eta$ Cha) but can be as
short as a few Myr.

\subsection{Binaries}

No binaries are present in our calculations, even though there are 
eight known or probable binary systems in the $\eta$ Cha cluster. 
In order to be able to compare the observations to the simulations, 
the masses of the binary components are combined to make a single 
star. The simulations thus contain the correct mass in stars but 
omit interactions due to single-binary and binary-binary scattering. 
In a forthcoming paper, we shall include an input binary population 
and undertake a detailed comparison with observations of the binary 
systems in $\eta$ Cha.

\begin{table*}[htbp]
  \centering
\begin{tabular}{|c|c|c|c|c|c|c|c|c|c|c|c|c|c|c|c|c|c|c|c|}
\hline
 $N_{init}$ & \multicolumn{7}{c|}{50} & \multicolumn{3}{c|}{30} &
 \multicolumn{3}{c|}{40} & \multicolumn{3}{c|}{60} &
 \multicolumn{3}{c|}{70} \\
\hline
$R_0$ (pc) & 0.3 & 0.1 & 0.05 & 0.03 & \multicolumn{2}{c|}{0.01} &
 0.005 & \multicolumn{2}{c|}{0.01} & 0.005 & \multicolumn{2}{c|}{0.01}
 & 0.005 & \multicolumn{2}{c|}{0.01} & 0.05 &
 \multicolumn{2}{c|}{0.01} & 0.05 \\
\hline
primordial &&&&&&&&&&&&&&&&&&& \\
mass segregation & \raisebox{0.9ex}{-} &
 \raisebox{0.9ex}{-} & \raisebox{0.9ex}{-} &
 \raisebox{0.9ex}{-} & \raisebox{0.9ex}{-} &
 \raisebox{0.9ex}{x} & \raisebox{0.9ex}{-} &
 \raisebox{0.9ex}{-} & \raisebox{0.9ex}{x} &
 \raisebox{0.9ex}{-} & \raisebox{0.9ex}{-} &
 \raisebox{0.9ex}{x} & \raisebox{0.9ex}{-} &
 \raisebox{0.9ex}{-} & \raisebox{0.9ex}{x} &
 \raisebox{0.9ex}{-} & \raisebox{0.9ex}{-} &
 \raisebox{0.9ex}{x} & \raisebox{0.9ex}{-} \\
\hline
\end{tabular}
  \caption{Summary of initial conditions for each simulation}
  \label{inicond}
\end{table*}

\subsection{Calculations}

For each set of initial conditions summarized in Table~\ref{inicond},
we made 100 realisations running an adapted version of the code NBODY3
written by Aarseth (1999). It is a direct, high precision N-body
integrator that uses the sophisticated Kustaanheimo-Stiefel (1965)
two-body regularization to treat close encounters. There is no
softening parameter. In total, 1900 simulations have been performed.

The stellar system containing initially $N_{init}$ sources in the mass
range $0.01-4 \msun$ is placed in the Galactic potential which defines 
a tidal radius 
\begin{equation*}
r_t=\left( \frac{G\,M_c}{4A(A-B)} \right)^{1/3}
\end{equation*}
where $A$ and $B$ are the Oort's constants, and $M_c$ is the cluster
mass. When the distance of an object to the cluster centre is larger
than $2 \times r_t$, it is considered to have escaped from the cluster
and it is removed from the calculation. Each calculation proceeds
until the number of bound systems falls below 14 or until the age
becomes larger than 10 Myr. The masses, positions and velocities of
the systems which remain bound to the cluster are output every $\sim
0.1$ Myr. There are about 100 outputs per simulation on average,
i.e. about 190000 output files were produced in total.


\section{Results and analysis}

The aim of the N-body simulations is to find initial conditions which
are able to reproduce the actual state of the $\eta$ Cha cluster, when
starting with a number of stellar systems $N_{init}$ imposed by a
log-normal IMF. Each output at each time step is analysed and
compared to the observations and several criteria have to be fulfilled
before it is considered to be a good match.

\subsection{Number of systems}
\label{section_ntot}

First, we look at the number of bound systems $N_{1pc}$ located within 1 
pc of the cluster centre in each output. If the number is larger than 22 
or smaller than 14, we do not consider the result as acceptable.

\begin{figure*}
\centering
\includegraphics[width=0.3\textwidth]{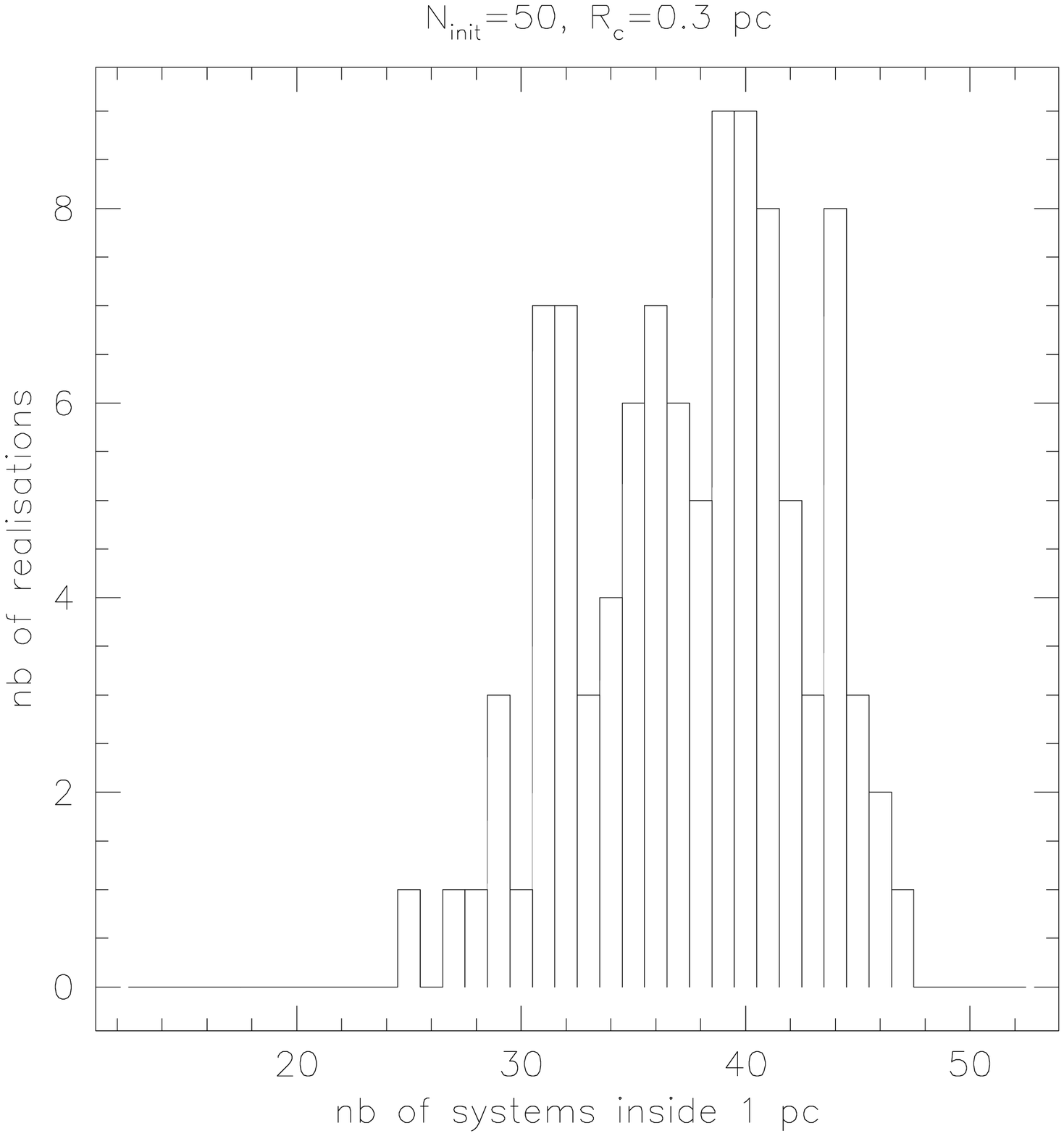}
\includegraphics[width=0.3\textwidth]{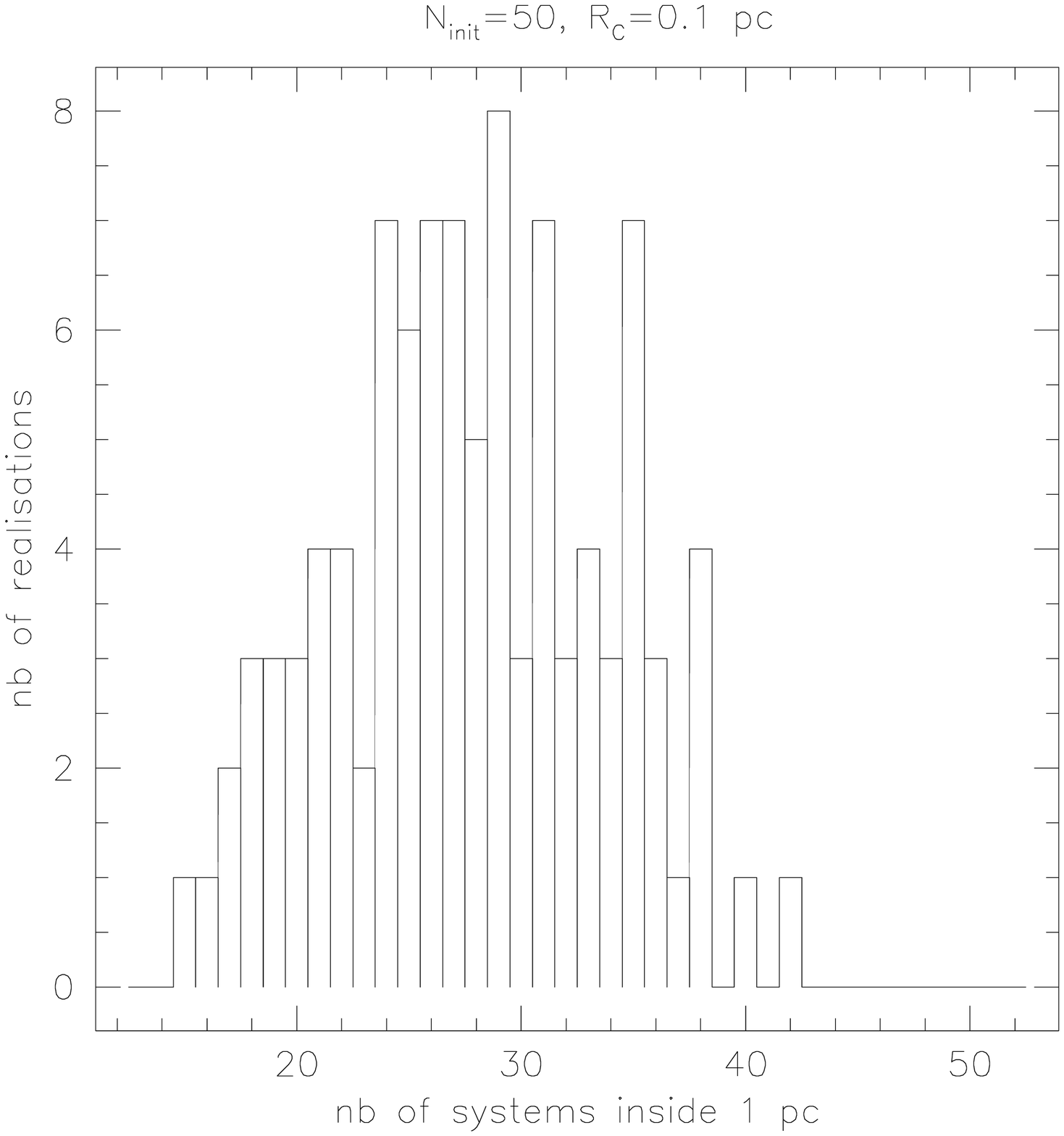}
\includegraphics[width=0.3\textwidth]{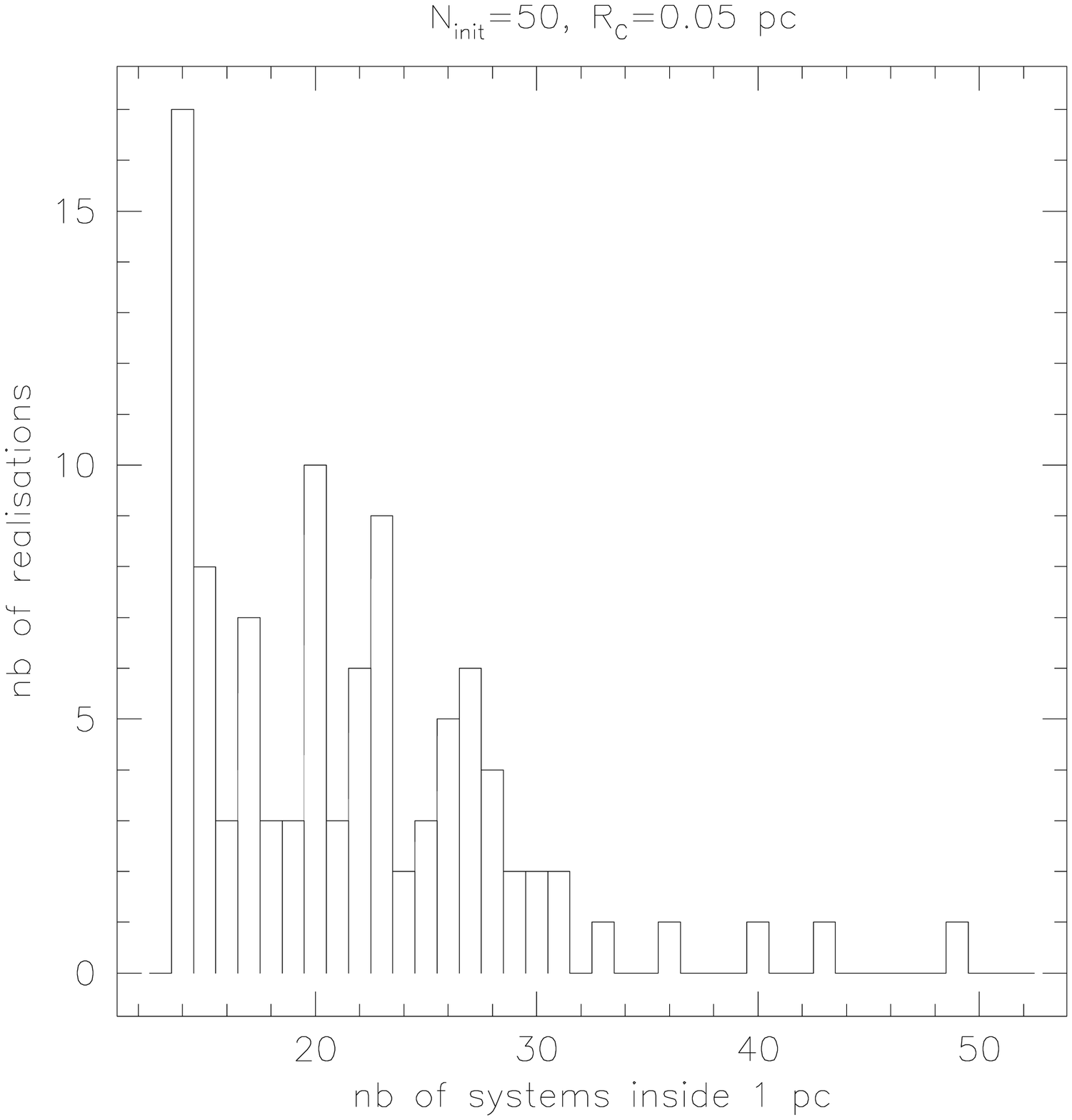}\\
\includegraphics[width=0.3\textwidth]{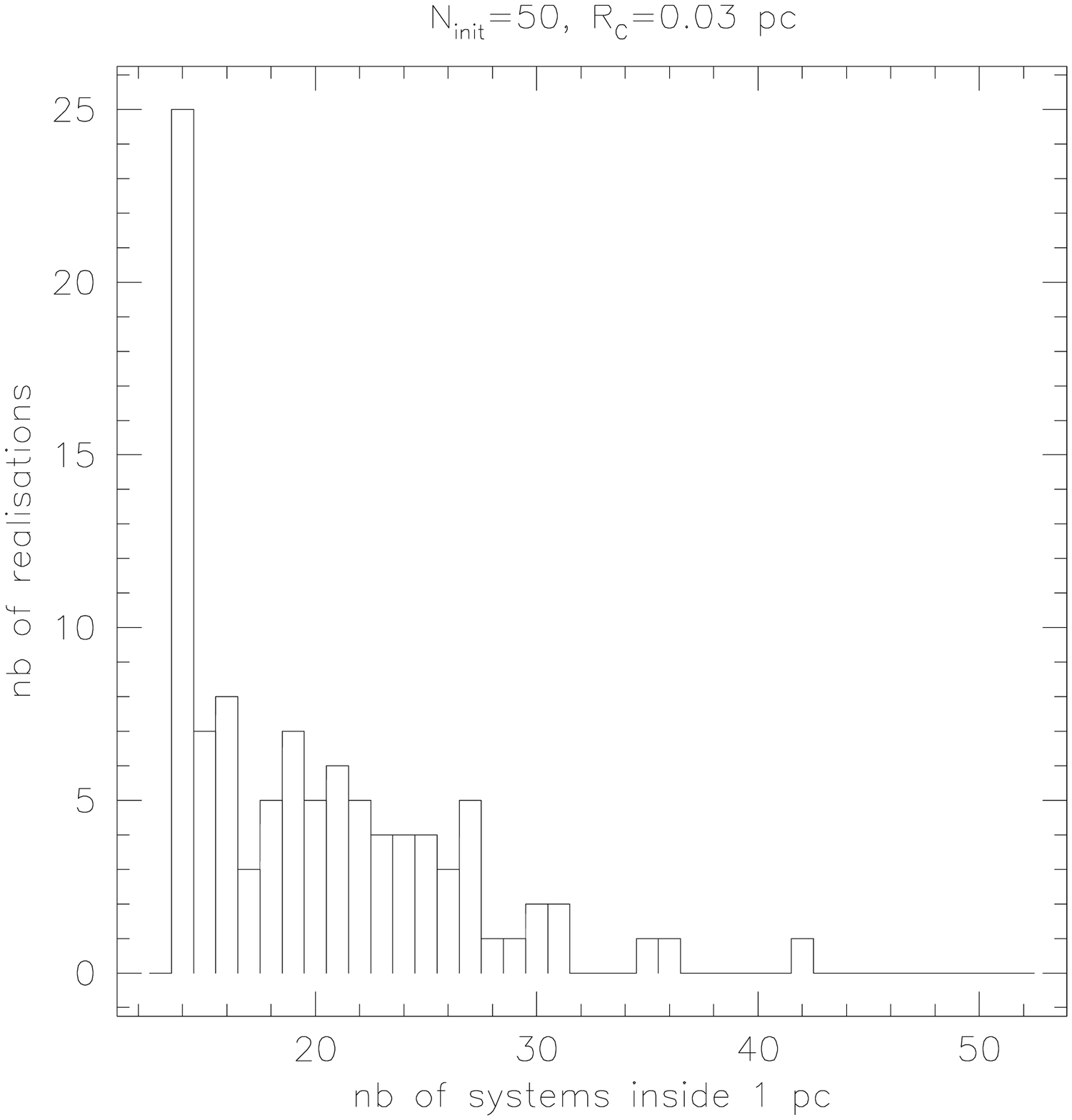}
\includegraphics[width=0.3\textwidth]{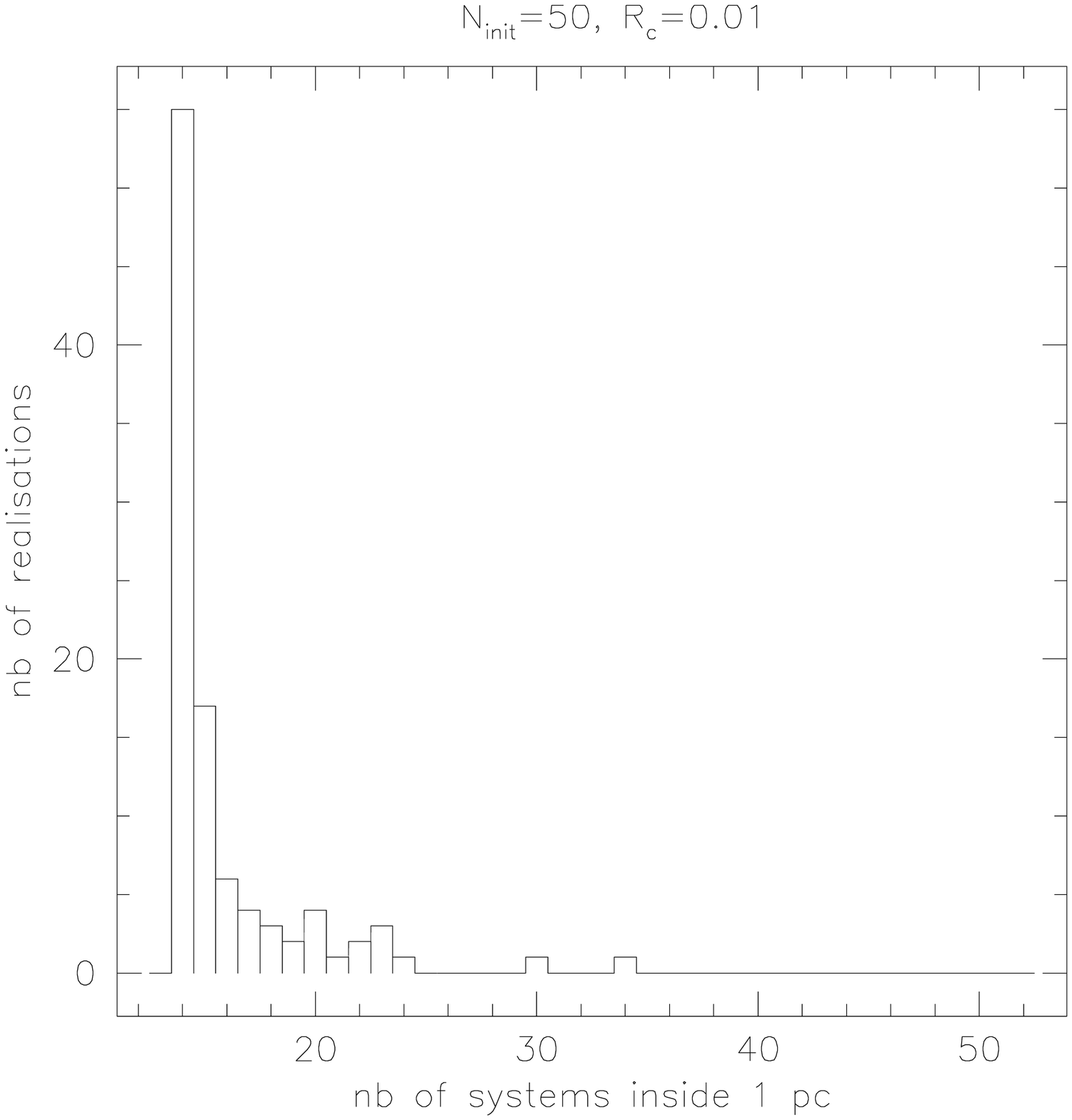}
\includegraphics[width=0.3\textwidth]{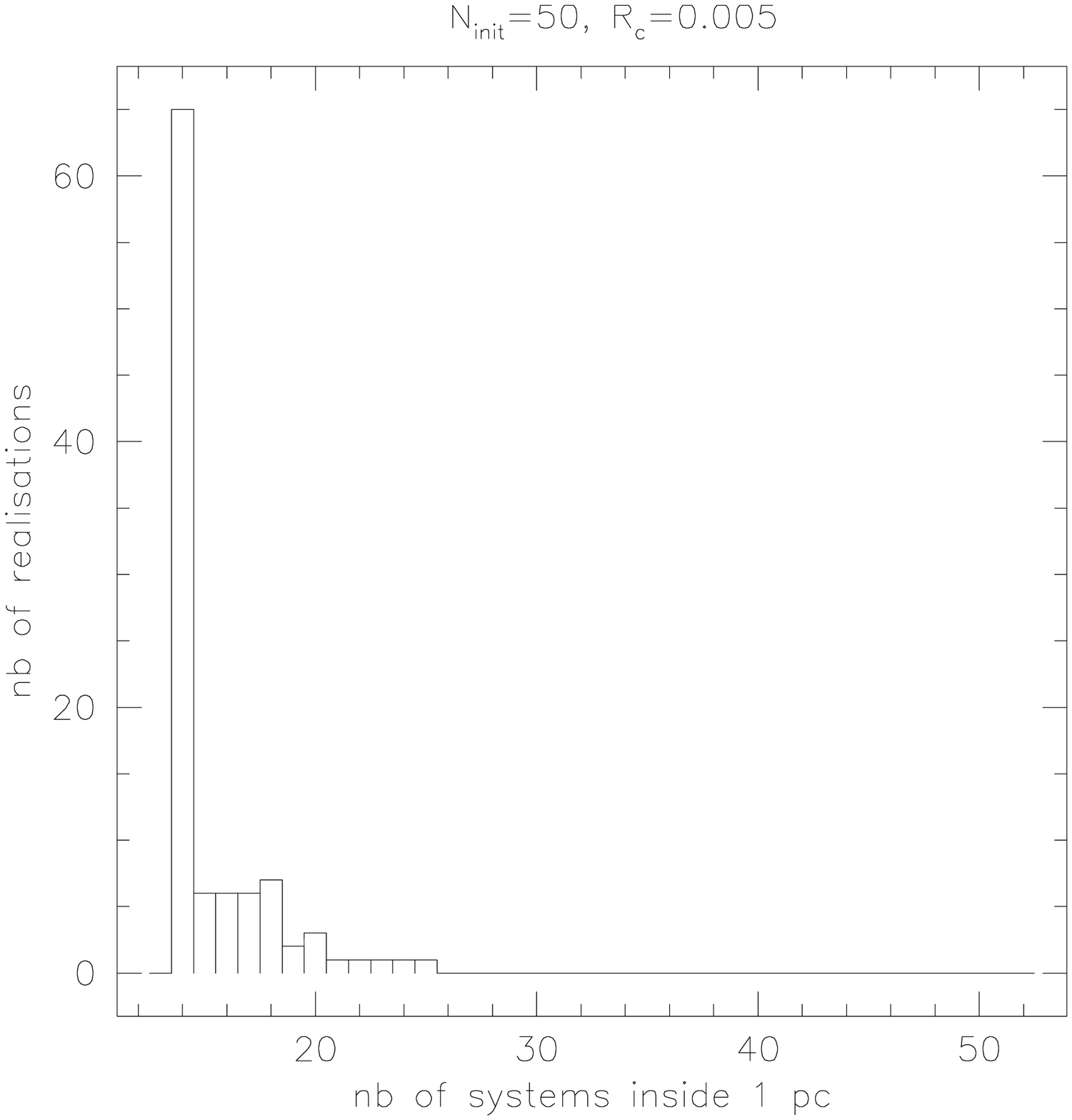}
\caption{Number of systems inside a radius of 1 pc for $N_{init}=50$
  and $R_0=0.3$, 0.1, 0.05, 0.03, 0.01 and 0.005 pc at the end of the
  calculation. This corresponds to an age of 10 Myr or less if the
  number of systems in the simulation falls below 14.}
\label{ntot}
\end{figure*}

We used this simple criterion alone for the first set of simulations 
starting with $N_{init}=50$ to estimate the initial cluster radius $R_0$ 
and the initial density required to lose the majority of the cluster 
members through dynamical interactions in less than 10 Myr. With $R_0=0.3$ 
pc, the cluster is too diffuse and the two-body encounters are not 
effective enough to eject members. Even after 10 Myr (the maximum duration 
of the simulations), more than half the systems ($> 25/50$) remained inside 
a radius of 1 pc for all of the realisations (see Fig.~\ref{ntot}). 

\begin{figure}[htbp]
\centering
\includegraphics[width=0.45\textwidth]{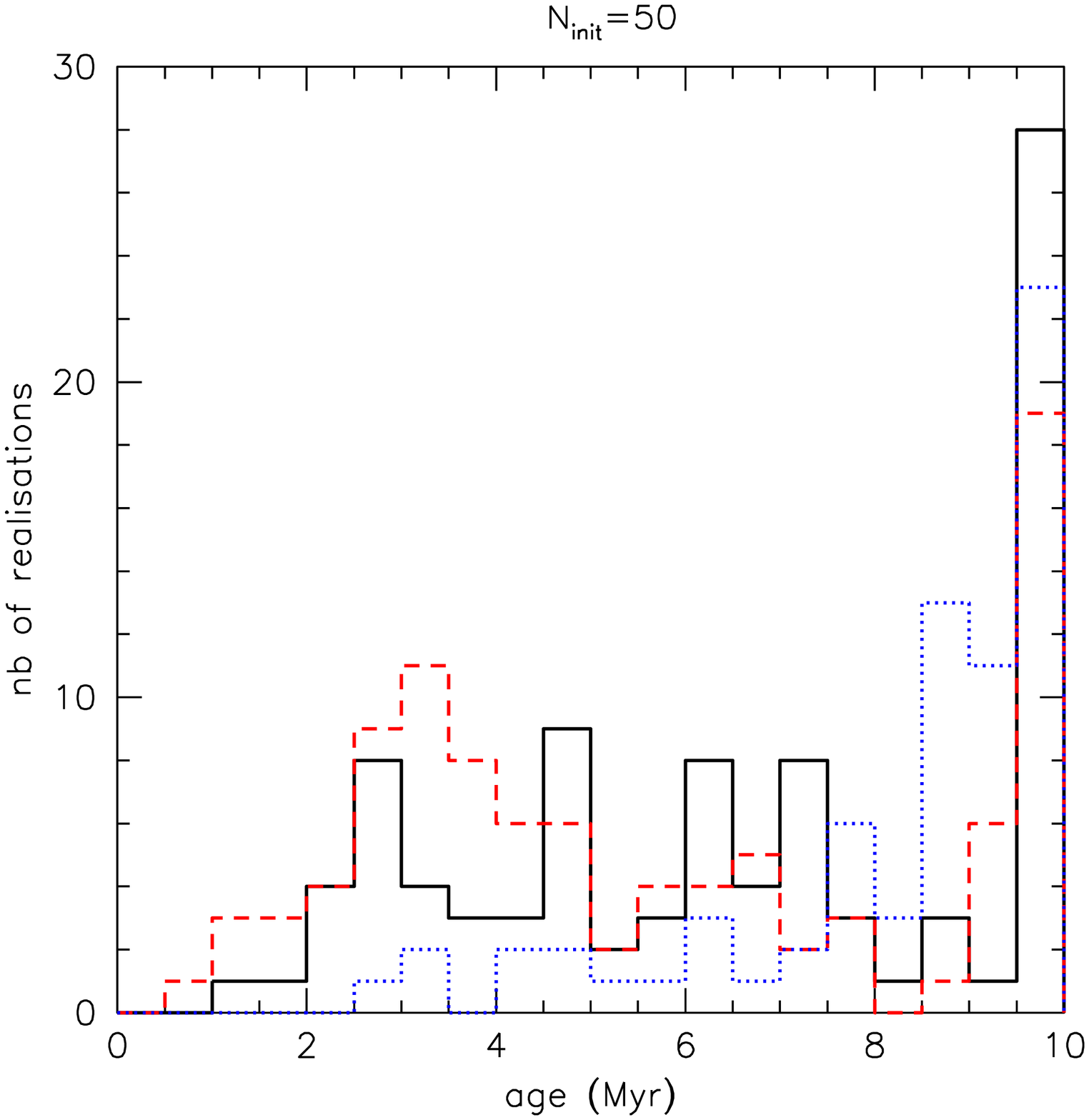}
\caption{``Age'' at the end of the simulations started with $N_{init}=50$
  and $R_0=0.03$ (dotted line), 0.01 (solid line) and 0.005 pc (dashed
  line). Here ``age'' means time at which $N_{1pc}$ dropped to $<14$ (if
  this is $<10$ Myr) or 10 Myr (if $N_{1pc}\le22$ at an age of 10
  Myr). The remaining simulations (for which $N_{1pc}>22$ at 10 Myr)
  are not shown here.} 
\label{age}
\end{figure}

The situation improves when we start the calculations with a more
compact configuration. As expected, more objects are found outside 1pc
and the cluster population spreads out more quickly. With
$N_{init}=50$ and $R_0=0.1$ pc initially, 21 realisations out of 100
end up with $N_{1pc}\le22$ but none of them reach $N_{1pc}=14$ in less
than 10 Myr. For $R_0=0.01$ pc and 0.005 pc however, most of the
calculations are stopped because they get to the final state with
$N_{1pc}=14$ before 10 Myr. We can see from Figure~\ref{age} that the
more compact the configuration, the more quickly $N_{1pc}=14$ is attained.
Bearing in mind that $\eta$ Cha has an age of $\sim 10$ Myr and that the
cluster lost its gas component probably after a few Myr (corresponding
to the time at which our N-body calculations start), the most
favorable age at the end of our simulations to reproduce $\eta$ Cha
should roughly be between 4 and 8 Myr.  The results from
Figure~\ref{age} suggest that the best initial cluster radius $R_0$ is
between 0.01 and 0.005 pc. We thus eliminated all the other values for
$R_0$ as initial conditions and we retained $R_0=0.005$ and 0.01 pc to
run the second set of simulations with $N_{init}=30$, 40, 50, 60 or 70
in order to find the best value of $N_{init}$ capable of matching the
observed cluster characteristics.

\begin{figure}
\centering
\includegraphics[width=0.24\textwidth]{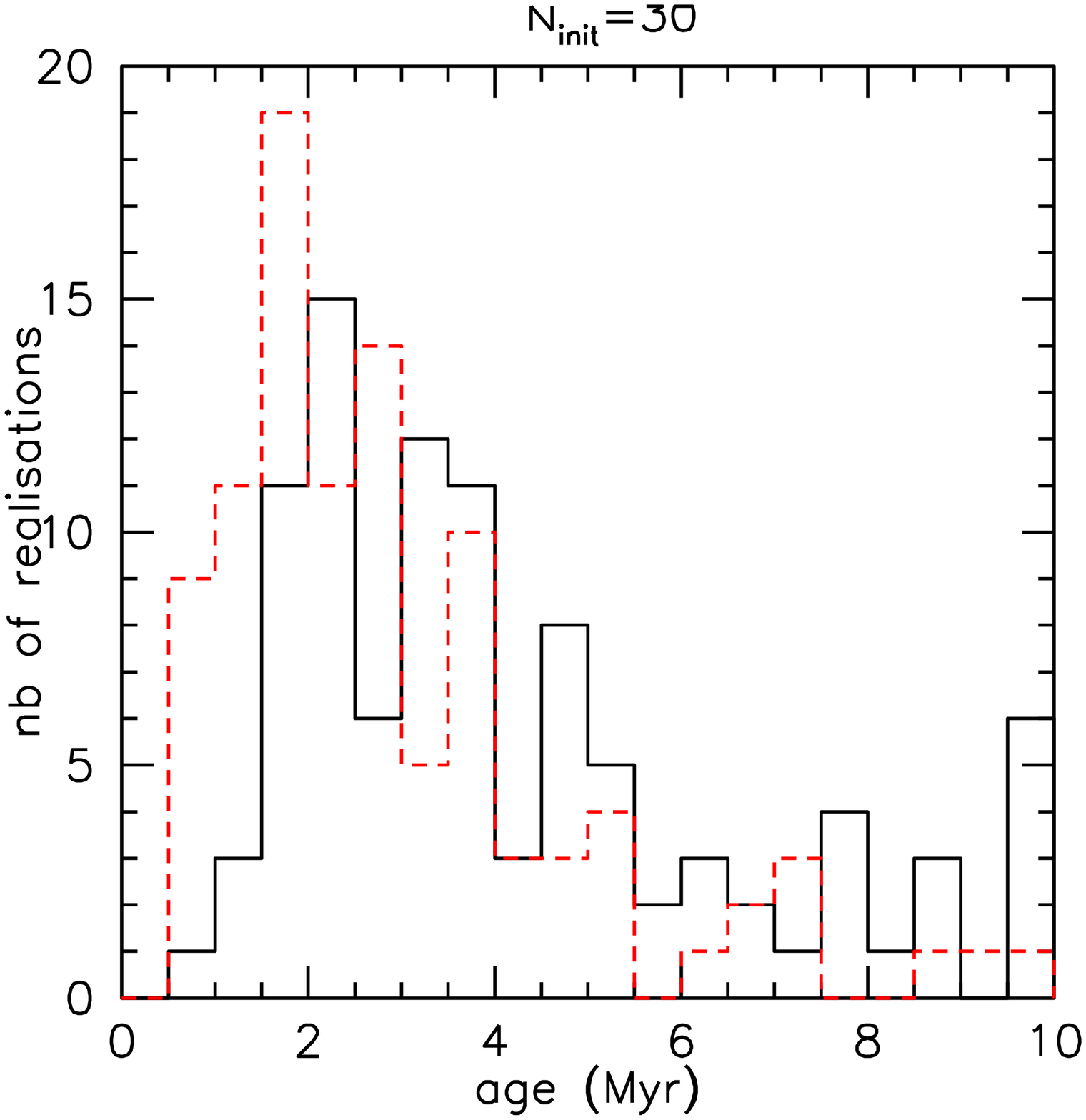}
\includegraphics[width=0.24\textwidth]{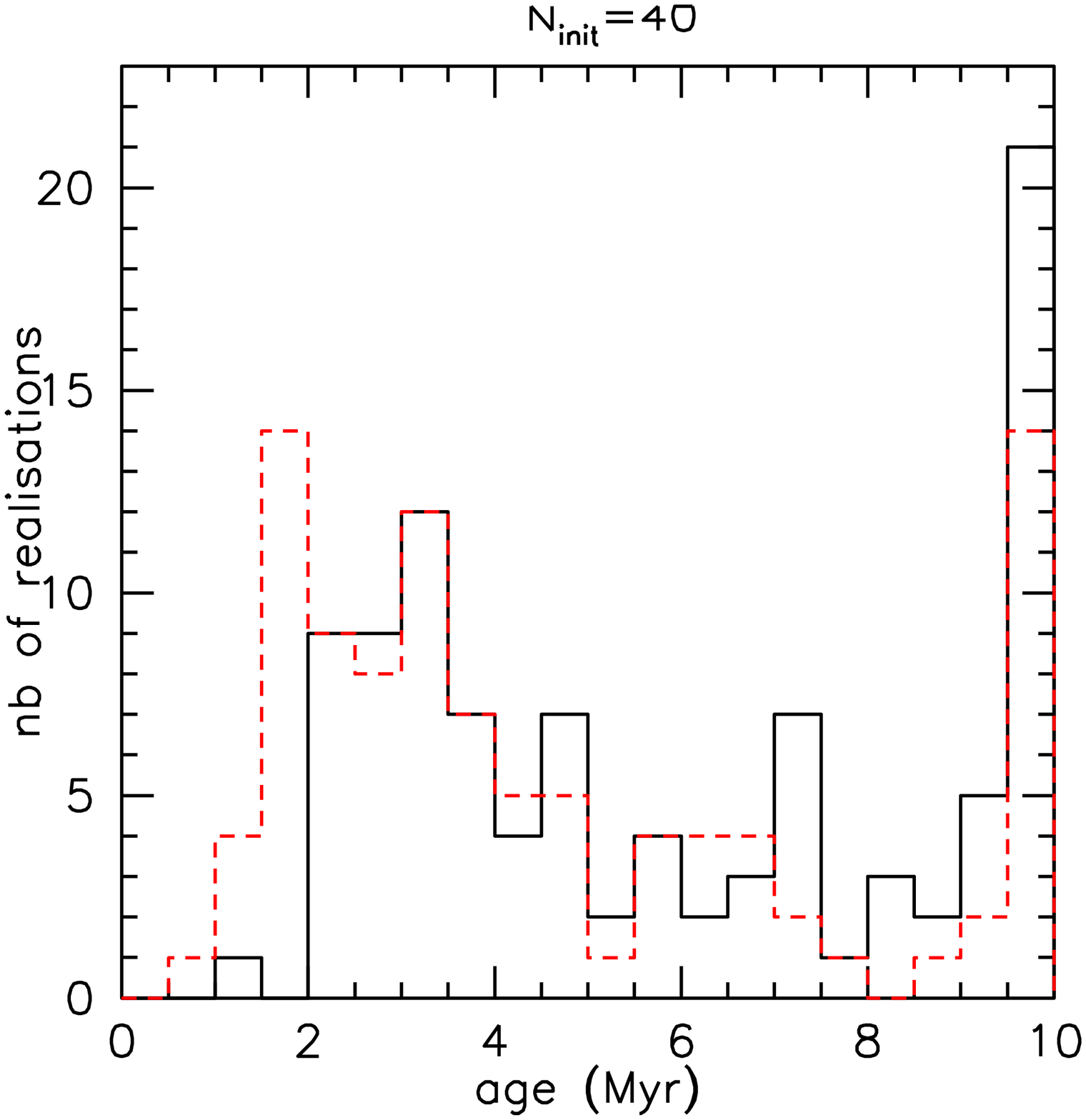}\\
\includegraphics[width=0.24\textwidth]{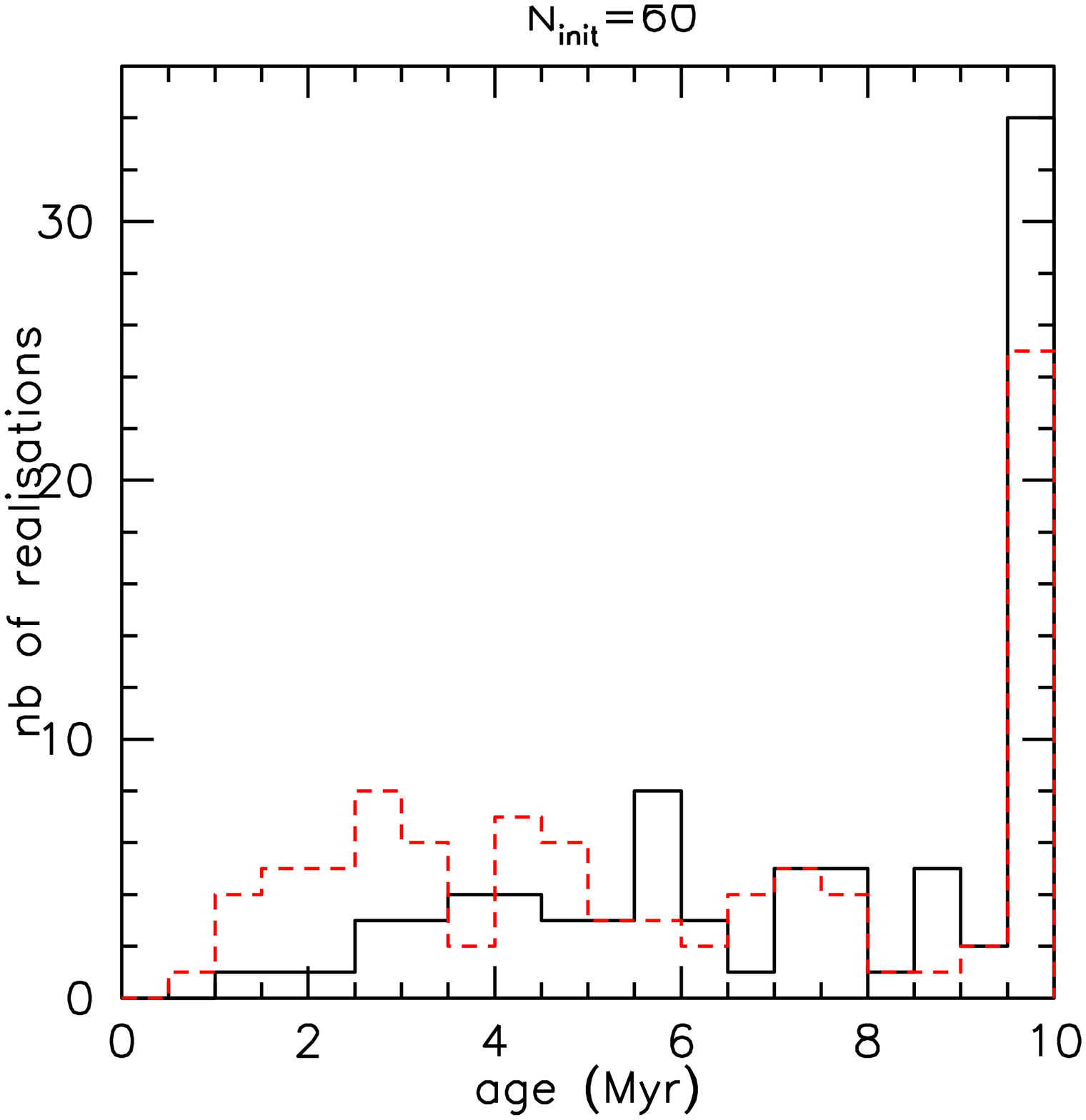}
\includegraphics[width=0.24\textwidth]{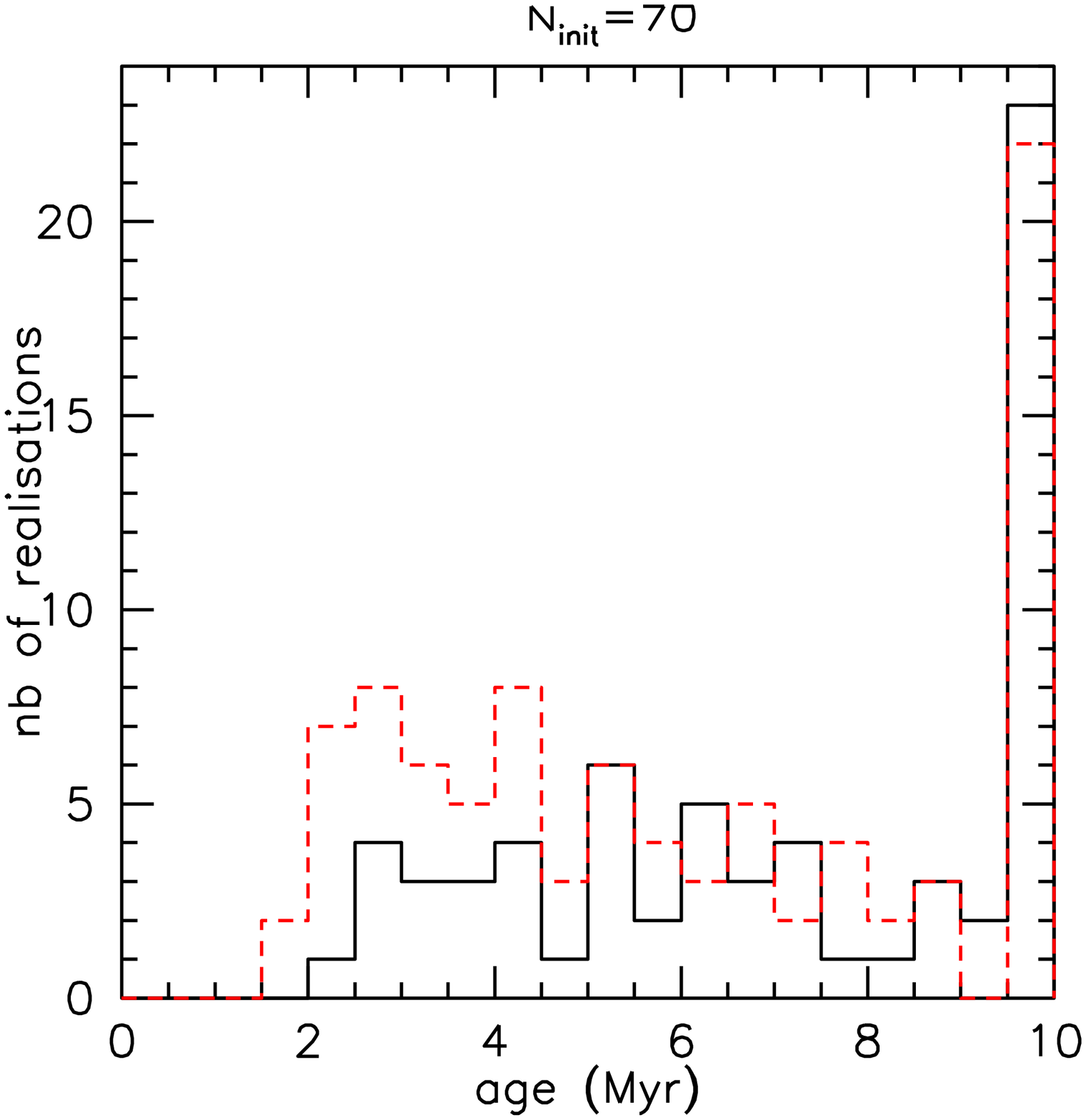}
\caption{Age at the end of the simulations started with $N_{init}=30$,
40, 60 and 70 which end up with $14\le N_{1pc}\le22$. In each panel
the solid histogram corresponds to $R_0=0.01$ pc and the dotted one to
0.005 pc.}
\label{age_ninit}
\end{figure}

By applying the same test on the results of these simulations, we
found that most of them reach $14\le N_{1pc}\le22$ before 10 Myr apart
from $N_{init}=70$ and $R_0=0.01$ pc where only 66 calculations fulfill
this criterion (see Table~\ref{nreal_ntot}). This is due to the fact
that there are more objects in the cluster and that it takes longer to
remove most of them from the centre. This indicates that initial
conditions with $N_{init}\ge 70$ are not favorable to reproduce $\eta$
Cha. Then we looked at the age distribution of the calculations which
ended with $14\le N_{1pc}\le22$ (Figure~\ref{age_ninit}). We found
that most clusters starting with $N_{init}=30$ lost their members very
quickly. There is a broad peak around $\sim2-3$ Myr in the distribution, 
which is probably too young to reproduce $\eta$ Cha. This suggests that 
simulations starting with $N_{init}\le30$ are less likely to match the 
observations. However, all of the other initial conditions with 
$N_{init}=40$, 50 or 60 and $R_0=0.01$ or 0.005 pc
seem to be a good match to the observed properties according to the
first criterion based solely on the number of objects left inside a
radius of 1 pc.

\begin{table}[htbp]
\centering
\begin{tabular}{|c|c|c|c|c|c|}
\hline
$N_{init}$ & 30 & 40 & 50 & 60 & 70 \\
\hline
$R_0=0.01$ pc & 97 & 99 & 94 & 87 & 66 \\
\hline
$R_0=0.005$ pc & 98 & 98 & 97 & 94 & 90 \\
\hline
\end{tabular}
\caption{Number of realisations out of 100 yielding $N_{1pc}\le22$
  within 10 Myr for $N_{init}=30$, 40, 50, 60, 70 and $R_0=0.01$ and
  0.005 pc.}
\label{nreal_ntot}
\end{table}

\subsection{Number of stars with $m\ge1\msun$}

When starting with a compact configuration ($R_0\le0.01$ pc), the
two-body interactions are strong enough to eject a large number of 
objects in a few Myr so that less than 22 stars remain inside a 
radius of 1 pc from the cluster centre. However, we do not want 
this to be the case for the most massive stars as four 
objects with a mass between 1 and $4\msun$ are observed in the 
cluster. Therefore the second criterion we used is to check that 
at least 3 stars more massive than 1.5$\msun$ are included in 
$N_{1pc}$, i.e. $N_{m\ge1.5\msun}\ge3$.

This second test allows us to eliminate more than half of the calculations 
that fulfill the first criterion, but does not really help us to define 
the best initial conditions as the trend is the same for all of them. In 
particular, the proportion of realisations which pass this test for a given 
$N_{init}$ is similar for $R_0=0.005$ or 0.01 pc (see Table~\ref{nreal_nmass}).

\begin{table}[htbp]
\centering
\begin{tabular}{|c|c|c|c|c|c|}
\hline
$N_{init}$ & 30 & 40 & 50 & 60 & 70 \\
\hline
$R_0=0.01$ pc & 33 & 51 & 44 & 45 & 39 \\
\hline
$R_0=0.005$ pc & 36 & 46 & 47 & 43 & 43 \\
\hline
\end{tabular}
\caption{Number of realisations out of 100 ending up with $N_{1pc}\le22$
  and $N_{m\ge1.5\msun}\ge3$ for $N_{init}=30$, 40, 50, 60, 70 and
  $R_0=0.01$ and 0.005 pc.} 
\label{nreal_nmass}
\end{table}

\subsection{Number of remaining brown dwarfs}

The third criterion is the comparison between the lower mass function
resulting from the simulations and the observations. No brown dwarf or
very low mass (VLM) star with $m\le 0.1\msun$ has been found within 1.5 
degrees of the cluster centre (Luhman 2004) and this has to be reproduced 
by the simulations. Therefore, we require that $N_{VLM}$, the number of 
objects with $m\le 0.1\msun$ inside 1.5 degrees, is 0 or 1 to consider 
a result as acceptable.

This is a very stringent constraint. We saw in Section
~\ref{section_ntot} that it is not too hard to eject a large number of
cluster members from the centre when starting with a very compact
configuration. Almost all the realisations with $R_0=0.01$ or 0.005 pc
end up with $14\le N_{1pc}\le22$.  It is already more difficult to
keep the most massive stars in the core -- only half of the
realisations passed this test -- but it is even harder to get all the
very low mass objects far away from the cluster centre in less than 10
Myr. Table~\ref{nreal_nbd} gives the number of realisations which
fulfill criteria 1 and 3 alone ($14\le N_{1pc}\le22$ and
$N_{VLM}\le1$) and all of criteria 1,2 and 3 ($14\le N_{1pc}\le22$,
$N_{m\ge1.5\msun}\ge3$ and $N_{VLM}\le1$). We can see that the number
drops down, around 10 or even less for some initial conditions, even
if we relax the second test on $N_{m\ge1.5\msun}$. This suggests in
particular that $R_0=0.01$ pc is not favorable to reproduce $\eta$
Cha, apart perhaps for $N_{init}=40$.

\begin{table}[ht]
\centering
\begin{tabular}{|c|c|c|c|c|c|c|c|c|c|c|}
\hline
$N_{init}$ & \multicolumn{2}{c|}{30} & \multicolumn{2}{c|}{40} &
  \multicolumn{2}{c|}{50} & \multicolumn{2}{c|}{60} &
  \multicolumn{2}{c|}{70} \\
\hline
$R_0=0.01$ pc & 10 & 2 & 8 & 8 & 6 & 0 & 1 & 0 & 2 & 1 \\
\hline
\raisebox{-0.5ex}{$R_0=0.01$ pc} &&&&&&&&&& \\
+ mass seg. & \raisebox{0.75ex}{2} & \raisebox{0.75ex}{0} &
\raisebox{0.75ex}{5} & \raisebox{0.75ex}{5} & \raisebox{0.75ex}{3} &
\raisebox{0.75ex}{2} & \raisebox{0.75ex}{4} & \raisebox{0.75ex}{0} &
\raisebox{0.75ex}{5} & \raisebox{0.75ex}{2} \\
\hline
$R_0=0.005$ pc & 12 & 12 & 12 & 10 & 4 & 2 & 10 & 8 & 11 & 2 \\
\hline
\end{tabular}
\caption{Number of realisations out of 100 ending with $14\le
  N_{1pc}\le22$ and $N_{VLM}\le1$ (first column) and with $14\le
  N_{1pc}\le22$, $N_{m\ge1.5\msun}\ge3$ and $N_{VLM}\le1$ (second
  column) for each initial conditions $N_{init}=30$, 40, 50, 60, 70
  and $R_0=0.01$ pc, 0.005 pc and 0.01 pc with primordial mass
  segregation.}
\label{nreal_nbd}
\end{table}

The reason why it is so difficult to fulfill the three criteria at the
same time is that the loss rate becomes roughly the same for all stars, 
independently of their mass, when the initial configuration is very 
compact. In order to keep the massive stars in the centre, the dynamical 
interactions should not be too strong to kick them out but then, even 
if some brown dwarfs escaped the cluster core, several still remain. 
In less than 10 Myr, the mass segregation favoring the evaporation of 
the lower mass objects does not really have time to occur. Simulations 
starting with a large number of systems $N_{init}$ like 60 or 70 have 
a larger number of brown dwarfs initially and it is very difficult to 
move them away in less than 10 Myr. Starting with $R_0=0.05$ pc instead 
of 0.01 pc helps to some extent as the configuration does not get too 
compact and the massive star are not removed from the core (which is 
the case for $N_{init}=70$ and $R_0=0.005$ pc).

In order to reproduce the mass distribution of $\eta$ Cha, one needs
to lose more brown dwarfs and VLM stars than higher mass stars. We can
therefore wonder if primordial mass segregation would help as it could
yield such a preferential escape rate. However, this does not seem to
be the case: the number of good realisations with $R_0=0.01$ pc does
not really get larger with primordial mass segregation
than without (see Table~\ref{nreal_nbd}). In some cases ($N_{init}=30$
and 40) it even gets smaller. This suggests that primordial mass
segregation is not required to reproduce $\eta$ Cha. In order to check
this result, we performed a last test on our simulations concerning
the radial distribution of the cluster members.

\subsection{Radial distribution}

We searched the ROSAT All-Sky Survey (RASS) for cluster members with
projected separations of up to 5 degrees. For 10 Myr-old stars T Tauri
stars at $\sim 100$ pc, the RASS is sensitive to mid-K to early-M
stars, i.e. to near-solar mass stars. To be candidate members,
RASS-selected objects should show high levels of lithium
indicative of young solar-type stars, photometry in agreement with the
placement of known cluster members in colour-magnitude diagrams, and
three-dimensional space motions consistent with ejection from the
cluster in the past $2-10$ Myr, at a velocity less than $\sim 5$
km/s. Only rarely is this set of quantities complete and available
with precision, e.g. optical photometry derived from plate scans is
poor and radial velocities are usually absent. Proper motions are
available from the UCAC2 catalogue, while a high-resolution
spectroscopic survey of RASS sources across the Chamaeleon region was
performed by Covino et al. (1997) which included a few known $\eta$
Cha stars. Within proper motion vectors $<10$ mas/yr of the cluster's
proper motion of $(\mu_{\alpha}, \mu_{\delta})= (-29.9, 27.5)$ mas/yr
(Mamajek et al. 1999), we found only a few additional RASS stars that
might meet these criteria for membership. Stars ejected from the
$\eta$ Cha with a velocity higher than 10 mas/yr, corresponding to 5
km/s at the distance of the cluster, will therefore not be sampled but
our simulations show that they are rare. High velocity objects are
ejected in the early stage of the cluster evolution and are already at
a much larger distance than 5 degrees from the cluster center. While
noting that these surveys are imprecise and/or incomplete, this
outcome indicates there is not a large number of undiscovered
solar-mass members within 5 degrees of the cluster core.

All of the cluster members are instead concentrated within 1 pc from the
cluster centre and only a few additional solar type stars, if any, can 
be located within 5 degrees. Therefore we require that the number of stars 
with $m\ge 0.5\msun$ located between 1 and 8.75 pc ($=5$ degrees for a
distance of 100 pc) from the cluster centre is less than 2. This constitutes 
our last criterion to check the ability of the simulations to reproduce 
the observations.

\begin{figure*}[htbp]
\centering
\includegraphics[width=0.9\textwidth]{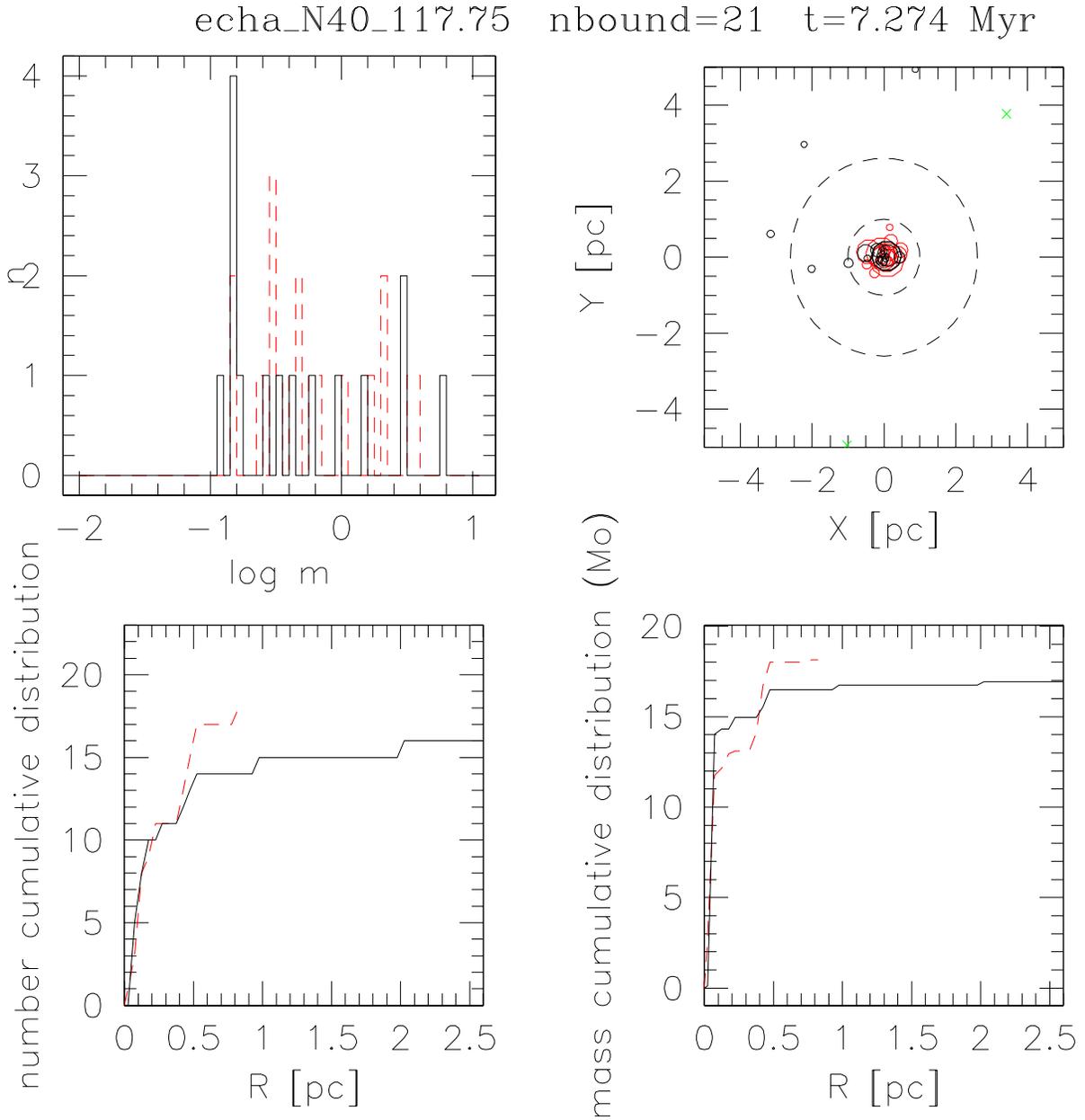}
\caption{Example of the result of a realisation considered as
  acceptable. The initial conditions are $N_{init}=40$ and $R_0=0.005$
  pc. Top left panel: cluster mass function from the simulation (solid
  histogram) and the observations (dashed histogram). Top right panel:
  Cluster member spatial distribution. The dashed circles centreed on
  the cluster correspond to a radius of 1 pc and 2.6 pc (=1.5 degrees
  at 100 pc). The stars with $m>0.1\msun$ are shown as circles and the
  symbols are larger for larger mass. The very low mass stars and
  brown dwarfs are shown as crosses. Bottom left panel: Cumulative
  distribution of the number of cluster members for the simulation
  (solid line) and the observations (dashed line). Bottom right panel:
  Mass cumulative distribution for the simulation (solid line) and the
  observations (dashed line).}
\label{example}
\end{figure*}

When performing this test on the realisations that already fulfill the 
three previous criteria, the number of remaining good results becomes 
very small for all the initial conditions (see Table~\ref{nreal_all}). 
Again we find that starting with primordial mass segregation can help 
in some cases but it is not necessary. A few percent of the simulations 
starting with $R_0=0.005$ pc can always reproduce $\eta$ Cha independently 
of $N_{init}$ which suggests that this corresponds to the best initial 
condition for the cluster radius. Similarly, $N_{init}=40$ is the most 
favorable initial number of systems, as we obtain good results for both 
$R_0=0.005$ and 0.01 pc. An example of a good realisation resulting from 
a calculation started with $N_{init}=40$ and $R_0=0.005$ is shown in
Figure~\ref{example}.

\begin{table}[ht]
\centering
\begin{tabular}{|c|c|c|c|c|c|}
\hline
$N_{init}$ & 30 & 40 & 50 & 60 & 70 \\
\hline
$R_0=0.01$ pc & 0 & 2 & 0 & 0 & 0 \\
\hline
\raisebox{-0.5ex}{$R_0=0.01$ pc} &&&&& \\
+ mass seg. & \raisebox{0.75ex}{0} & \raisebox{0.75ex}{2} &
\raisebox{0.75ex}{1} & \raisebox{0.75ex}{2} & \raisebox{0.75ex}{0} \\
\hline
$R_0=0.005$ pc & 4 & 5 & 2 & 2 & 1 \\
\hline
\end{tabular}
\caption{Number of realisations out of 100 fulfilling the full set of
  selection criteria for each initial conditions $N_{init}=30$, 40,
  50, 60, 70 and $R_0=0.01$ pc without and with primordial mass
  segregation and $R_0=0.005$ pc.}
\label{nreal_all}
\end{table}

It is important to note, that even if only a few percent of the
realisations fulfill all the above criteria, they remain valid across
several time steps, i.e. the time for which we have a good match with
the observations last for more than 1 Myr, typically at an age between
4 and 10 Myr. For example, in the realisation from which
Fig.~\ref{example} is drawn, the simulated cluster remains in
agreement with the observations between 3.5 and 10 Myr. The fact that
such a dynamical state is not transient indicates that the
calculations can mimic an observational state and that our simulations
can effectively reproduce $\eta$ Cha. The difference between the age
given by the simulation and the cluster age would then correspond to
the time at which the gas was lost.


\section{Conclusion and discussion}

\subsection{Summary of results}

To summarize, the criteria we used to consider that a result is
consistent with the observations are the following:
\begin{itemize}
\item the number of objects within 1 pc of the cluster centre is
  $14\le N_{1pc} \le 22$ at some point at $t<10$ Myr,
\item it includes at least 3 stars more massive than $1.5\msun$,
\item there is no more than one brown dwarf or VLM star with $m\le
  0.1\msun$ within 1.5 degrees,
\item and less than 2 stars with $m\ge 0.5\msun$ are located between 1
  and 8.75 pc (corresponding to 5 degrees at a distance of 100 pc).
\end{itemize}

We tested initial conditions with $R_0=0.3$ pc down to 0.005 pc and
$N_{init}=30$, 40, 50, 60 and 70 distributed according to a log-normal
IMF. None of the simulations starting with $R_0\ge0.01$ pc give good
results, except if $R_0=0.01$ pc and $N_{init}=40$. For the
calculations with $R_0=0.005$ pc however, there are always a few
realisations that match the observations whatever the value of
$N_{init}$.

We thus found that it is effectively possible to reproduce $\eta$ Cha,
which suggests in particular that the deficit of brown dwarfs and very
low mass stars observed in the present day cluster mass function may
not be due to a peculiar IMF but to dynamical evolution.

\subsection{Compactness of the core}

The most favorable initial condition in reproducing $\eta$ Cha is
$N_{init}=40$ and $R_0=0.005$ pc, which corresponds to an initial
central density of $3.7 \times 10^8$ stars/pc$^3$. In the example
shown Fig.\ref{example}, the total cluster mass at the beginning of
the calculation is $21.4\msun$. This corresponds to an initial
crossing time of 0.003 Myr which confirms a posteriori our assumption
that the relaxation phase after the gas expulsion is very short ($\le
0.1$ Myr). According to the maximum stellar mass -- cluster mass
relation discussed by Weidner \& Kroupa (2006), the mass of the
embedded cluster at the origin of $\eta$ Cha was of the order of
$30-40\msun$. This suggests that $30-40\%$ of the stars had been lost
when the simulation starts and that the initial density may have been
even higher than the value we obtain. Further calculations including
the treatment of gas disruption (as well as an input binary
population) will be done in a forthcoming paper to investigate this
effect.

Nevertheless, a density of $\sim 10^8$ stars/pc$^3$ is
extraordinarily high and potentially makes $\eta$ Cha a unique region
where star formation would have occurred in a very compact
configuration. It is 5 orders of magnitude larger than the initial
density taken by Kroupa \& Bouvier (2003) to simulate the aggregates
in Taurus and $2-3$ orders of magnitude larger than that used for the
ONC and Pleiades models (Kroupa et al. 2001). In the simulations
of massive star cluster formation performed by Bonnell et al. (2003),
the stellar density attains a maximum of $10^7$ to $10^8$ stars/pc$^3$
but this occurs only locally. At the cluster's scale the density is
typically $10^5$ stars/pc$^3$, 3 orders of magnitude smaller than what
we have. Observationally, such a high density has never been found in
any embedded cluster but this may be due to a resolution effect. An
initial radius $R_0=0.005$ pc corresponds to a half mass radius of
0.0035 pc or 7.4'' on the sky assuming a cluster distance of 97 pc. At
the distances to the youngest large-scale regions of star formation
such as the Carina nebula at $d=2.3$ kpc, such a cluster would not be
spatially resolved (or only marginally) by radio and IR observations
and would instead be seen as a bright single source. High angular
resolution observations (e.g. with ALMA) are needed to know if such
clusters exist.

\subsection{Binarity}

For the large majority of systems, the smallest nearest-neighbour
distance is obtained at $t=0$ and is around $5\times 10^{-4}-10^{-3}$
pc, which corresponds to $100-200$ AU. This is consistent with the
typical ejection velocities of a few km/s (Fig.~\ref{esc}) and
suggests that binaries cannot have separations larger than a factor of
a few times less than this distance. Kroupa \& Burkert (2001)
performed N-body calculations of extremely compact clusters, with a
central density similar to what we find for $\eta$ Cha. They start
with a binary proportion $f_b=1$ and a uniform logarithmic period
distribution in the range $4.5 \le \log P \le 5.5$, with $P$ in
days. This corresponds to a separation interval $\sim 20-100$ AU for
an equal mass binary of a total mass $1 \msun$. For an initial density
$\log \rho_{c}=8.9$ stars/pc$^3$, the binary proportion in this period
range falls down to $f_b=0.2$ very quickly, in a timescale of the
order of a crossing time. This is consistent with the results from
Brandeker et al. (2006) who found an upper limit of 18\% of wide
binaries at projected separations $>30$ AU, and also with the apparent
absence of circumbinary disks in {\it Spitzer} IRS spectra of $\eta$
Cha (Bouwman et al. 2006).

\subsection{Where are the escapers ?}

\begin{figure*}[htbp]
\centering
\includegraphics[width=0.9\textwidth]{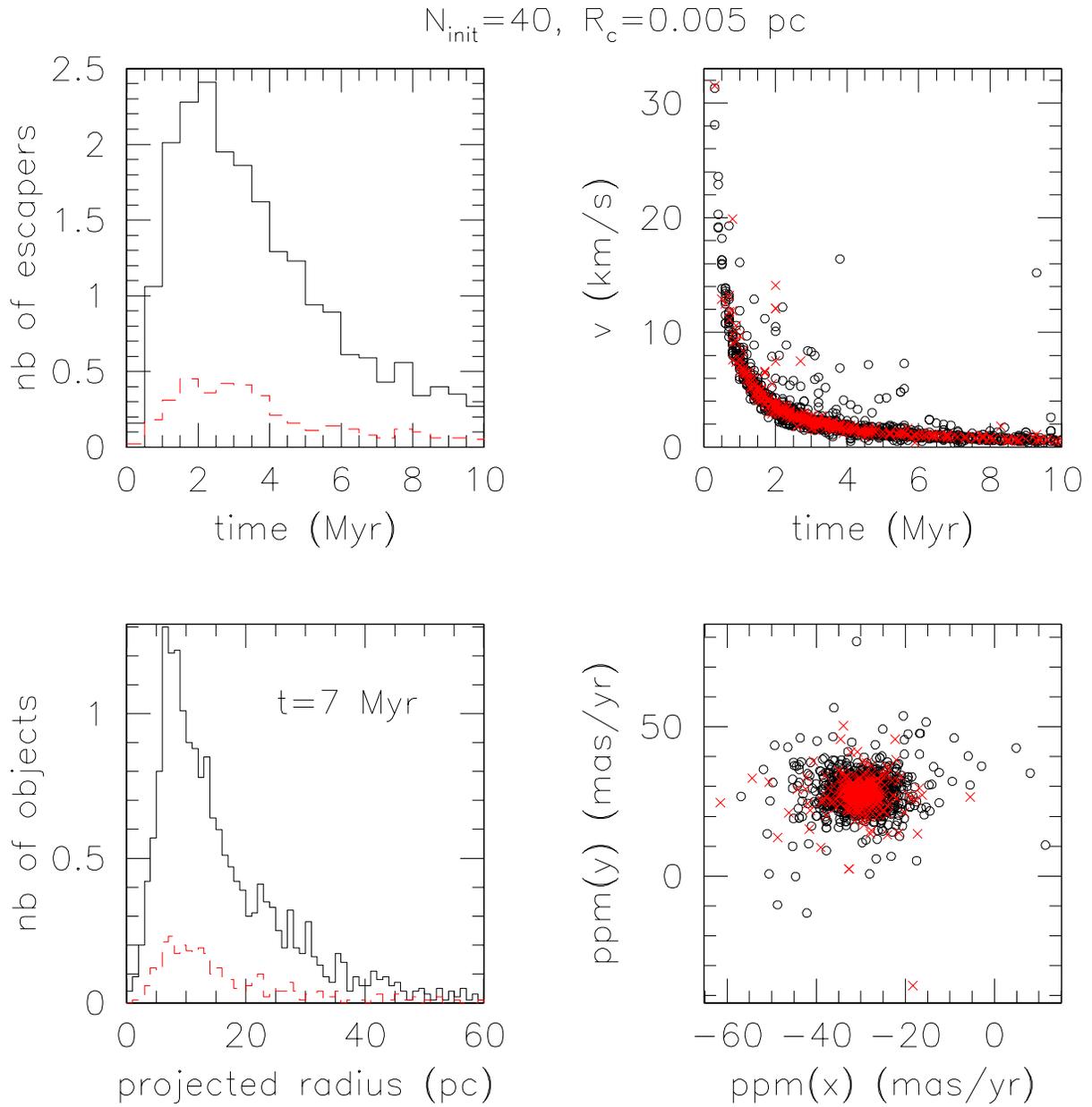}
\caption{Statistics over all the 100 simulations starting with
 $N_{init}=40$ and $R_0=0.005$ pc. Top left panel: Average number of
 escaped systems per time bin (0.5 Myr) with $m\ge 0.1 \msun$ (solid
 line) and $m\le 0.1 \msun$ (dashed line) as a function of time. Top
 right panel: Ejection velocity of all the escapers versus time (for
 all the 100 realisations). Crosses are VLM stars and brown dwarfs
 with $m\le 0.1 \msun$. Bottom left panel: Radial distribution of the
 escapers at t=7 Myr for $m\ge 0.1 \msun$ (solid line) and $m\le 0.1
 \msun$ (dashed line). The number of objects is averaged over the 100
 realisations and is given per radius bin of 1 pc. Bottom right panel:
 Proper motion diagram of all the escapers at t=7 Myr.}
\label{esc}
\end{figure*}

The objects that reach twice the tidal radius, i.e. $r=6$ to 8 pc
depending on the simulations, escape the cluster and are removed from
the calculations. But it is still possible to follow their evolution
afterwards because we know their position, velocity and the time at
which they escape. We simply assume that their movement is linear. The
velocity of the escapers is decreasing with time as the density gets
smaller and the interactions are softer. We find that it goes roughly
with $1/$time and that it is independent of the object mass, i.e. the
ratio of VLM objects to normal stars is roughly constant with radius
and is consistent with the input IMF. All the systems that are ejected
in the early phase (the first Myr) have a velocity of $\sim6$ km/s or
more which means that they are now more than $\sim$60 pc (or 30
degrees) away, given the age and distance of $\eta$ Cha. Most of the
objects are ejected between 1 and 4 Myr with a velocity in the range
$1-5$ km/s. If we assume that the simulations run for 7 Myr to
reproduce the cluster as in the example shown in Fig.~\ref{example},
they will be at a distance of at least 9 pc. The systems that are
ejected later have a smaller velocity and remain closer to the cluster
centre as there is less time to spread them away. Statistically, for
all the simulations starting with $N_{init}=40$ and $R_0=0.005$ pc, we
find a larger number of systems around $6-10$ pc in projected distance
(or $3.5-6$ degrees for a cluster distance of 97 pc) whatever the time
is (see Figure~\ref{esc}). The surface density is so small ($\sim
0.02$ objects/pc$^2$) that it would be quite difficult to observe this
population photometrically. However proper motion surveys might be
able to select for ejected cluster members as their projected
velocities should be distributed around the cluster motion of ($-29.9,
27.5$) mas/yr, which is quite high. Recently Cruz et al. (2007)
discovered a population of young ($5-50$ Myr) brown dwarfs in the
field. A few of them could come from $\eta$ Cha and would have been
ejected early with a high velocity ($>6$ km/s).

\begin{acknowledgements}

The authors wish to thank S.~Aarseth for allowing us access to his
N-body codes.

WAL acknowledges support from UNSW@ADFA SRG and FRG grants, and visitor 
programmes at LAOG and IoA. We are also grateful for assistance from 
the Australian Research Council-sponsored Australia-France Cooperation 
Fund in Astronomy (AFCOP).

\end{acknowledgements}



\end{document}